\documentstyle[12pt,epsfig]{article}     
                                               
\newlength{\dinwidth}                                               
\newlength{\dinmargin}                                               
\def\lapproxeq{\lower .7ex\hbox{$\;\stackrel{\textstyle                                               
<}{\sim}\;$}}                                               
\def\gapproxeq{\lower .7ex\hbox{$\;\stackrel{\textstyle                                               
>}{\sim}\;$}}                                               
\def\be{\begin{equation}}                                               
\def\ee{\end{equation}}                                               
\def\bea{\begin{eqnarray}}                                               
\def\eea{\end{eqnarray}}

\begin{document}                                               
\titlepage                                               
\begin{flushright}                                               
DTP/99/100 \\                                               
November 1999 \\                                               
\end{flushright}                                               
                                               
\vspace*{2cm}                                               
                                               
\begin{center}                                               
{\Large \bf Unintegrated parton distributions and prompt photon hadroproduction}                                               
                                               
\vspace*{1cm}                                               
M.A. Kimber$^a$, A.D. Martin$^a$ and M.G. Ryskin$^{a,b}$ \\                                               
                                              
\vspace*{0.5cm}                                               
$^a$ Department of Physics, University of Durham, Durham, DH1 3LE \\                                              
$^b$ Petersburg Nuclear Physics Institute, Gatchina, St.~Petersburg, 188350, Russia      
\end{center}                                               
                                               
\vspace*{1cm}      
      
\begin{abstract}                                               
We introduce a general expression which enables the parton distribution, unintegrated over    
the parton transverse momentum, to be obtained from the conventional parton densities.  We    
use the formalism to study the effects of the transverse momentum $q_t$ of the incoming    
partonic system on the calculation of the transverse momentum spectra of prompt photons    
produced in high energy $pp$ and $p\bar{p}$ collisions.  For the purposes of illustration, we    
use the double logarithm approximation.  For large $q_t$ we calculate the effect directly from    
the perturbative formalism, whereas for small $q_t$ we bound the effect using two extreme      
hypotheses.  In both $q_t$ domains we find that the shapes of the prompt photon spectra are      
not significantly modified, although the cross sections are enhanced.     
\end{abstract}                                              
                                      
\medskip      
\section{Introduction}      
     
The cross sections for hard hadronic processes are conventionally described in terms of      
universal parton distributions $a (x, \mu^2)$, with $a = xq$ or $xg$, convoluted with the     
cross sections of the partonic subprocesses calculated in perturbative QCD at some large scale     
$\mu$ characteristic of the subprocess.  For example, $\mu$ may be the transverse     
momentum of an outgoing parton or the mass of a heavy quark should it participate in the     
subprocess.  In this paper we present the formalism which enables the parton distributions     
$f_a (x, k_t^2, \mu^2)$, {\it unintegrated} over the parton transverse momentum $k_t$, to be     
constructed from the conventional integrated distributions $a (x, \lambda^2)$.  We will see     
that it is important to distinguish the scale $\lambda$ from the hard scale $\mu$.  We first     
give the general prescription, and then we take the limits which correspond to the leading    
$\log (1/x)$ BFKL approach \cite{BFKL} and to the double $\log$ DDT prescription    
\cite{DDT}.     
     
These unintegrated parton densities $f_a (x, k_t^2, \mu^2)$ should be used to describe any      
hard had\-ronic process, such as heavy quark production or the production of large $E_T$ jets
\cite{KT}.       
The only exception is deep inelastic scattering (DIS).  If the $k_t$ integration is performed in      
the convolution of the unintegrated parton distributions with the deep inelastic partonic      
subprocess then the remaining $x$ convolution is in terms of just the conventional parton      
distributions\footnote{A somewhat analogous example is the DIS renormalization scheme,      
where the higher order corrections are present in any process except DIS, which is used as the      
reference process.}.     
     
To demonstrate the use of the formalism we study a particularly relevant and topical process,      
namely prompt photon hadroproduction.  The production of prompt photons in high energy      
$pp$ collisions has long played a key role in      
constraining the gluon distribution of the proton at large $x$ through the dominance of the      
subprocess $gq \rightarrow \gamma q$.  The reason is that the gluon enters at leading order,      
unlike its contribution to the description of deep inelastic scattering.  However the description      
of the transverse momentum spectrum of the produced photons is more problematic than {\it      
inclusive} deep inelastic scattering.  The observed $p_{t \gamma}$ spectrum in $pp      
\rightarrow \gamma X$ (or $p\bar{p} \rightarrow \gamma X$) is found to be steeper than the      
prediction of perturbative QCD \cite{HUSTON,AUR}.  The explanation of this discrepancy      
is usually attributed to the intrinsic transverse momenta $k_t$ of the incoming partons, which      
are usually assumed to have a Gaussian-like $k_t$ distribution \cite{HUS,MRST}.       
Thus part of the observed $p_{t \gamma}$ comes from the initial partonic $k_t$ such that the      
hard subprocess singularity $d\hat{\sigma}/d\hat{t} \sim 1/p_{t \gamma}^4$ is approached      
more closely, and hence leads to a steeper $p_{t \gamma}$ spectrum.  However in order to      
describe the observed spectra it is necessary to introduce      
a $k_t$ spectrum with an average value which increases from $\langle k_t \rangle \sim      
0.5$~GeV to more than 2~GeV \cite{HUS} as the collision energy $\sqrt{s}$ increases from  
UA6, E706 \cite{UA6,E706} to Tevatron \cite{FNAL} energies.  Such large partonic  
$\langle k_t \rangle$ cannot originate solely from the large distance confinement domain, but  
must also have a significant perturbative QCD component.  Indeed it is easy to show that  
perturbative QCD\footnote{Another correction to the perturbative form of the $p_{t  
\gamma}$ spectrum comes from the resummation of $\log (1 - x_T)$ terms, where $x_T  
\equiv 2p_{t \gamma}/\sqrt{s}$ \cite{CATANI}.  This effect only changes the spectrum at  
the larger values of $x_T$, close to the kinematic boundary.} gives a non-zero $k_t$ with a  
distribution in which $\langle k_t \rangle$ increases as $\sqrt{s}$.  Ref.~\cite{LL} also  
contains a discussion of the partonic $k_t$ generated by perturbative QCD, but comparison  
with the $p_{t \gamma}$ data is based on a phenomenological Gaussian $k_t$ distribution.     
     
Here we study the role of both the non-perturbative and perturbative components of the      
incoming partonic $k_t$ in describing the observed $p_{t \gamma}$ spectrum in high energy      
prompt photon hadroproduction, that is $pp$ (or $p\bar{p}) \rightarrow \gamma X$.       
Perturbative QCD is applied, at leading order, in the domain of {\it large} partonic transverse      
momentum $k_t > q_0$, where $q_0 \sim 1-2$~GeV is the starting point of DGLAP      
evolution in the global parton analyses \cite{MRST,CTEQ}.  In the {\it small} $k_t < q_0$      
region it proves sufficient to consider two extreme hypotheses for the normalised    
non-perturbative or intrinsic $k_t$ spectrum:     
\begin{itemize}     
\item[(a)] a $\delta (k_t^2)$ distribution, i.e. no intrinsic $k_t$ from the confinement region,     
\item[(b)] a wide Gaussian distribution of width $\langle k_t^2 \rangle \sim \langle q_0^2      
\rangle$.     
\end{itemize} From
these extreme possibilities we conclude that the intrinsic partonic transverse momentum      
has a rather small effect on the $p_{t \gamma}$ spectrum.  Moreover we find the effects of      
$k_t$ from the perturbative domain $(k_t > q_0)$ do not significantly change the $p_{t      
\gamma}$ spectrum.     
     
In Section 2 we introduce the unintegrated parton distributions and the parton-parton     
luminosity function ${\cal L}_{ab} (x_1, x_2, q_t)$, which is the probability of finding the     
incoming $a, b$ partons (with momenta specified by $x_1, k_{1t}^2$ and $x_2, k_{2t}^2$)     
with net transverse momentum $\mbox{\boldmath $q$}_t = \mbox{\boldmath $k$}_{t1} +     
\mbox{\boldmath $k$}_{t2}$.  In Section 3 we apply the formalism to the production of  
prompt photons in high energy $pp$ or $p\bar{p}$ collisions.  We    
convolute the luminosity functions with the hard $gq \rightarrow \gamma q$ and $q\bar{q}    
\rightarrow \gamma g$ subprocess cross sections, taking into account the boost of the final    
$\gamma +$parton system coming from the incoming partonic $q_t$.  In Section 4, we show    
the effect of both the perturbative and non-perturbative components of $q_t$ on the    
predictions for the $p_{t \gamma}$ distributions at different $\sqrt{s}$, together with the    
measured $p_{t \gamma}$ spectra.  Finally in Section 5 we present our conclusions.     
     
\section{Unintegrated partons and the luminosity function}      
      
To explore the effects of partonic transverse momenta we need to know the distribution of a       
parton $a$, say as a function of its transverse momentum $k_t$, as well as of $x$.  A       
straightforward way to obtain such a distribution is to consider the DGLAP       
evolution\footnote{For the $g \rightarrow gg$ splitting we need to insert a factor $z^\prime$       
in the last integral of (\ref{eq:a1}) to account for the identity of the produced gluons.}      
\be      
\label{eq:a1}      
\frac{\partial a}{\partial \ln \lambda^2} \; = \; \frac{\alpha_S}{2 \pi}       
\: \left [ \int_x^{1 - \Delta} \: P_{aa^\prime} (z) \: a^\prime \left (\frac{x}{z}, \lambda^2      
\right ) \: dz \: - \: a (x, \lambda^2) \: \sum_{a^\prime} \: \int_0^{1 - \Delta} \:      
P_{a^\prime a} (z^\prime) \,dz^\prime \right ],     
\ee      
where the (integrated) parton density, $a (x, \lambda^2)$, denotes $xg (x, \lambda^2)$ or $xq      
(x, \lambda^2)$.  The first term on the right-hand-side describes the number of partons $\delta      
a$ emitted in the interval $\lambda^2 < k_t^2 < \lambda^2 + \delta \lambda^2$, while the      
second (virtual) term reflects the fact that the parton $a$ disappears after the splitting.  The      
second contribution may be resummed to give the survival probability $T_a$ that the parton      
$a$ with transverse momentum $k_t$ remains untouched in the evolution up to the      
factorization scale.  The survival probability is given by the double logarithmic Sudakov      
factor \cite{MW}     
\be      
\label{eq:a2}      
T_a (k_t, \mu) \; = \; \exp \left (- \: \int_{k_t^2}^{\mu^2} \: \frac{\alpha_S (p_t)}{2 \pi} \:       
\frac{dp_t^2}{p_t^2} \: \sum_{a^\prime} \: \int_0^{1 - \Delta} \: P_{a^\prime a} (z^\prime)     
\: dz^\prime \right ).   
\ee      
Thus the probability to find a parton $a$ with transverse momentum $k_t$ which initiates our    
hard process, with factorization scale $\mu$, is      
\be      
\label{eq:a3}      
f_a (x, k_t^2, \mu^2) \; = \; \left ( \frac{\alpha_S (k_t)}{2 \pi} \: \int_x^{1 - \Delta} \:       
P_{aa^\prime} (z) \: a^\prime \left (\frac{x}{z}, k_t^2 \right ) \: dz \right ) \: T_a      
(k_t, \mu).      
\ee      
Now we have to specify the value of the infrared cut-off $\Delta$, which is introduced to     
protect the $1/(1-z^\prime)$ singularity in the splitting functions arising from soft gluon    
emission.  In the original     
DGLAP equation (\ref{eq:a1}) for integrated partons this singularity is cancelled between the     
real and virtual contributions.  However after the resummation of the virtual terms the real     
soft gluon emission has to be accounted for explicitly as it changes the transverse momentum     
of the parton.  That is we have to find the physically appropriate choice of the cut-off     
$\Delta$, which comes from the coherence effect \cite{MW}.  The most convenient way is to     
go to the Breit frame where the angular ordering condition becomes the requirement that no     
soft gluons are emitted in the backward direction.  In this frame the parton energy $E =     
\mu$.  Then the integral over the soft gluon momentum $dz^\prime/(1-z^\prime) =     
d\omega/\omega$ covers the interval $k_t < \omega < E$, where $\omega$ is the energy of     
the soft gluon.  That is we take $\Delta = k_t/E = k_t/\mu$.  Of course the same $\Delta$     
must be used in both the real emission integral in (\ref{eq:a3}) and in the survival probability     
$T$ in (\ref{eq:a2}).  Below we consider two limits of the general expression (\ref{eq:a3})    
for the unintegrated parton density.   
    
In the leading $\log (1/x)$ or BFKL limit the virtual contribution in the DGLAP equation,     
(\ref{eq:a1}), is neglected and the survival probability $T = 1$.  Hence we come back to the     
familiar prescription    
\be    
\label{eq:b3}    
f_a (x, k_t^2) \; = \; \left . \frac{\partial (a (x, \lambda^2))}{\partial \ln \lambda^2} \right     
|_{\lambda^2 = k_t^2}.   
\ee    
Note that in this limit the unintegrated parton density does not depend on $\mu$, since there     
are no $\log (1/x)$ terms in the scale dependent part.    
    
In the double logarithm limit, (\ref{eq:a3}) may be written    
\be    
\label{eq:c3}    
f_a (x, k_t^2, \mu^2) \; = \; \left . \frac{\partial}{\partial \ln \lambda^2} \: \left [ a (x,     
\lambda^2) \: T_a (\lambda, \mu) \right ] \right |_{\lambda = k_t}    
\ee    
with survival probability    
\be    
\label{eq:d4}    
T_a (\lambda, \mu) \; = \; \exp \left (- \int_{\lambda^2}^{\mu^2} \: \frac{\alpha_S     
(p_t)}{2\pi} \: \frac{dp_t^2}{p_t^2} \: \int_0^{1 - \Delta} \: \frac{dz^\prime}{1 - z^\prime} \:     
2 C_a \right )    
\ee    
where $C_q = C_F = (N_C^2 - 1)/2 N_C$ and $C_g = C_A = N_C$.  In this limit the cut-off     
$\Delta$ is numerically small so one can choose the value of $\Delta$ in (\ref{eq:a1}) equal     
to the value of $\Delta$ in (\ref{eq:d4}).  Therefore the second term of the right-hand-side of     
(\ref{eq:a1}) cancels the derivative $\partial T_a/\partial \ln \lambda^2$.  To be precise, the  
only double logarithmic contribution comes from the singular $1/(1 - z^\prime)$ part of the  
diagonal 
splitting function $P_{aa} (z^\prime)$ in (\ref{eq:a1}) which exactly equals the last integrand  
$2 C_a/(1 - z^\prime)$ in (\ref{eq:d4}).      
      
To obtain the parton-parton luminosity function we have to perform a convolution over the       
distributions of the two incoming partons $a, b$      
\be 
\label{eq:xx} 
{\cal L}_{ab} (x_1, x_2, q_t) \; = \; \int \: f_a (x_1, k_1^2, \mu^2) \: f_b (x_2, k_2^2, \mu^2)       
\: \delta^{(2)} (\mbox{\boldmath $k$}_1 + \mbox{\boldmath $k$}_2 - \mbox{\boldmath      
$q$}_t) \: \frac{d^2 k_1 d^2 k_2}{\pi k_1^2 k_2^2},      
\ee 
where, for simplicity, we have omitted the $t$ subscript on the $k_i$.  The luminosity    
element, ${\cal L}_{ab} dq_t^2$, gives the probability that the incoming partons have a net    
square transverse momentum in the interval $(q_t^2, q_t^2 + dq_t^2)$.  To leading order all       
the transverse momenta are strongly ordered, and so either $k_1 \ll k_2 \simeq q_t$ or $k_2       
\ll k_1 \simeq q_t$.  In the first case we may integrate over $k_1$ giving      
\be      
\label{eq:a5}      
\int^{q_t} \: f_a f_b \: \frac{d^2 k_1}{\pi k_1^2} \; = \; \left [a (x_1, q_t^2) \: T_a (q_t, \mu)      
\right ] \: f_b (x_2,       
q_t^2, \mu^2),      
\ee      
and vice versa for $k_2 \ll k_1$.  Thus the sum of both contributions may be expressed in the       
compact form      
\be      
\label{eq:a6}      
{\cal L}_{ab} (x_1, x_2, q_t) \; = \; \left . \frac{\partial}{\partial \lambda^2} \: \left [a       
(x_1, \lambda^2) \: T_a (\lambda, \mu) \: b (x_2, \lambda^2) \: T_b (\lambda, \mu) \right ]       
\right |_{\lambda = q_t}.      
\ee      
A similar expression for the Drell-Yan process was originally obtained in the classic work of       
ref.~\cite{DDT}.     
     
\section{The prompt photon $p_{t \gamma}$ distribution}      
      
\begin{figure}\centering
\includegraphics{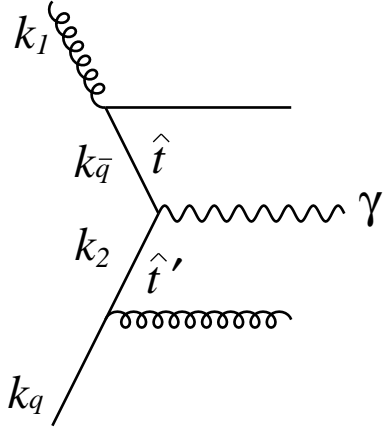}
\caption{A schematic diagram describing both the subprocesses $gq \rightarrow \gamma      
q$ and $\bar{q}q \rightarrow \gamma g$, in which the hard momentum transfer squared is      
either $\hat{t}$ or $\hat{t}^\prime$ respectively.}
\label{fig1}
\end{figure}

To obtain the cross section for inclusive prompt photon hadroproduction we convolute the       
hard subprocess cross sections for $gq \rightarrow \gamma q$ etc.~with the parton luminosity       
of (\ref{eq:a6}).  We obtain      
\bea      
\label{eq:a7}      
E_\gamma \: \frac{d \sigma}{d^3 p_\gamma} & = & \frac{d \sigma}{d \eta_\gamma \pi d      
p_{t \gamma}^2} \; = \; \int \: {\cal L} (x_1, x_2, q_t) \: \frac{dx_1}{x_1} \:      
\frac{dx_2}{x_2} \: dq_t^2 \nonumber \\      
& & \nonumber \\      
& & \times \; \frac{d \hat{\sigma}}{d\hat{t}} \: d\hat{t} \: \frac{d \phi}{2 \pi} \: \delta       
(\eta_\gamma - \ldots) \: \delta (p_{t \gamma}^2 - \ldots) \: \theta \left ( \mu^2 -       
q_t^2 \right ) \: \theta (| \hat{t} | - q_t^2)       
\eea      
where $\phi$ is the azimuthal angle between $\mbox{\boldmath $q$}_t$ and the photon       
transverse momentum $\mbox{\boldmath $p$}_t^\prime$ in the hard subprocess.  That is      
$p_t^\prime = \frac{1}{2} M \sin \theta$, where $\theta$ is the polar scattering angle in the      
hard subprocess and $M^2 = \hat{s}$ is the invariant mass squared of the produced $\gamma      
q$ system.  We thus have     
\be     
\label{eq:a8}     
\hat{t} \; = \; - \: \frac{M^2}{2} \; (1 - \cos \theta)  
\ee     
in the hard subprocess cross section $d\hat{\sigma}/d\hat{t}$.  The theta functions in  
(\ref{eq:a7}) impose the physical requirements that the partonic      
transverse momentum should be less than, first, the scale $\mu$ and, second, the momenta in      
the hard subprocess.  The last condition comes from the fact that at leading order all the      
transverse momenta along the chain are strongly ordered.  For the hard subprocess initiated by      
the gluon $(k_1)$ and the quark $(k_2)$ of Fig.~\ref{fig1}
 the virtualities $k_1^2, k_2^2 \ll | \hat{t}      
|$.  It means that $q_t^2 \simeq {\rm max} (k_1^2, k_2^2) < | \hat{t} |$.  Fig.~\ref{fig1} also      
describes the $\bar{q}q \rightarrow \gamma g$ subprocess but now $k_2^2 = \hat{t}^\prime$      
plays the role of the hard momentum transfer and strong ordering implies $k_{\bar{q}}^2,      
k_q^2 \ll | \hat{t}^\prime |$.  Thus the product ${\cal L} d \hat{\sigma}/d\hat{t}$ in      
(\ref{eq:a7}) should be understood as the sum over the different subprocesses, that is, as  
$\sum {\cal L}_{ab} d \hat{\sigma}_{ab}/d\hat{t}_{ab}$ where $\hat{t}_{gq} = \hat{t}$  
and $\hat{t}_{\bar{q}q} = \hat{t}^\prime$ in Fig.~\ref{fig1}.     
     
The introduction of the delta functions in (\ref{eq:a7}) is simply a technical device to enable      
the differential cross section to be expressed in terms of $\eta_\gamma$ and $p_{t      
\gamma}^2$.  Let us study the relevant kinematics.  If the outgoing $\gamma +$ parton      
system, of mass $M$      
and rapidity $\eta_M$, is boosted with the incoming \lq\lq partonic\rq\rq~transverse      
momentum $q_t$, then the rapidity $\eta_\gamma$ and transverse momentum $p_{t      
\gamma}$ of the photon become     
\bea     
\label{eq:a9}     
\eta_\gamma & = & \eta_M \: + \: \frac{1}{2} \ln \: \left (\frac{M_\perp + q_s + M \cos      
\theta}{M_\perp + q_s - M \cos \theta} \right ), \\     
& & \nonumber \\     
\label{eq:a10}     
p_{t \gamma}^2 & = & \frac{M^2}{4} \: \sin^2 \theta \: + \: \frac{(q_t^2 + q_s^2)}{4} \: + \:      
\frac{M_\perp q_s}{2},     
\eea     
where     
\be     
\label{eq:a11}     
q_s \; \equiv \; q_t \: \sin \theta \: \sin \phi, \quad\quad M_\perp^2 \; \equiv \; M^2 + q_t^2.     
\ee     
Equations (\ref{eq:a9}) and (\ref{eq:a10}) specify the missing parts of the delta functions in      
(\ref{eq:a7}).  As the variables $\eta_M$ and $M$ have been used to specify the Lorentz      
boost it is convenient to write the integration over the luminosity in (\ref{eq:a7}) in terms of      
$dM^2 d\eta_M dq_t^2$.  This is most easily done by noting that     
\be     
\label{eq:a12}     
\int \frac{dx_1}{x_1} \: \int \frac{dx_2}{x_2} \: \delta (M^2 - x_1 x_2 s + q_t^2) \: \delta      
\left (\eta_M - {\textstyle \frac{1}{2}} \ln (x_1/x_2) \right ) \; = \; \frac{1}{M_\perp^2}.     
\ee     
We may use the first delta function in (\ref{eq:a7}) to perform the $d \eta_M$ integration and      
the second delta function $\delta (p_{t \gamma}^2 - \ldots$) to perform the $M^2$      
integration.  Then, using $d | \hat{t} | = \frac{1}{2} M^2 \sin \theta d \theta$, we can rewrite      
(\ref{eq:a7}) in the form     
\be     
\label{eq:a13}     
E_\gamma \: \frac{d \sigma}{d^3 p_\gamma} \; = \; \int {\cal L} (M^2, \eta_M, q_t^2) \:      
\frac{dq_t^2}{M_\perp^2} \: \frac{d \phi}{2 \pi} \: \frac{d \hat{\sigma}}{d | \hat{t} |}      
\: \frac{2 M^2 d \theta}{\sin \theta \: + \: (q_t/M_\perp) \sin \phi} \; \theta \left (\mu^2 \: - \:      
q_t^2 \right ) \: \theta \left ( | \hat{t} | \: - \: q_t^2 \right ),     
\ee     
with $M_\perp$ specified by $\delta (p_{t \gamma}^2 - \ldots)$, that is by     
\be     
\label{eq:a14}     
M_\perp \; = \; -q_t \frac{\sin \phi}{\sin \theta} \; \pm \; \frac{1}{\sin \theta} \: \sqrt{4 p_{t      
\gamma}^2 - q_t^2 \cos^2 \theta \cos^2 \phi}.     
\ee     
Of course we may only use solutions of this latter equation which satisfy $M_\perp > q_t$.     
     
\section{Effects of partonic $q_t$ on the photon spectrum}     
     
\begin{figure}\centering
\includegraphics{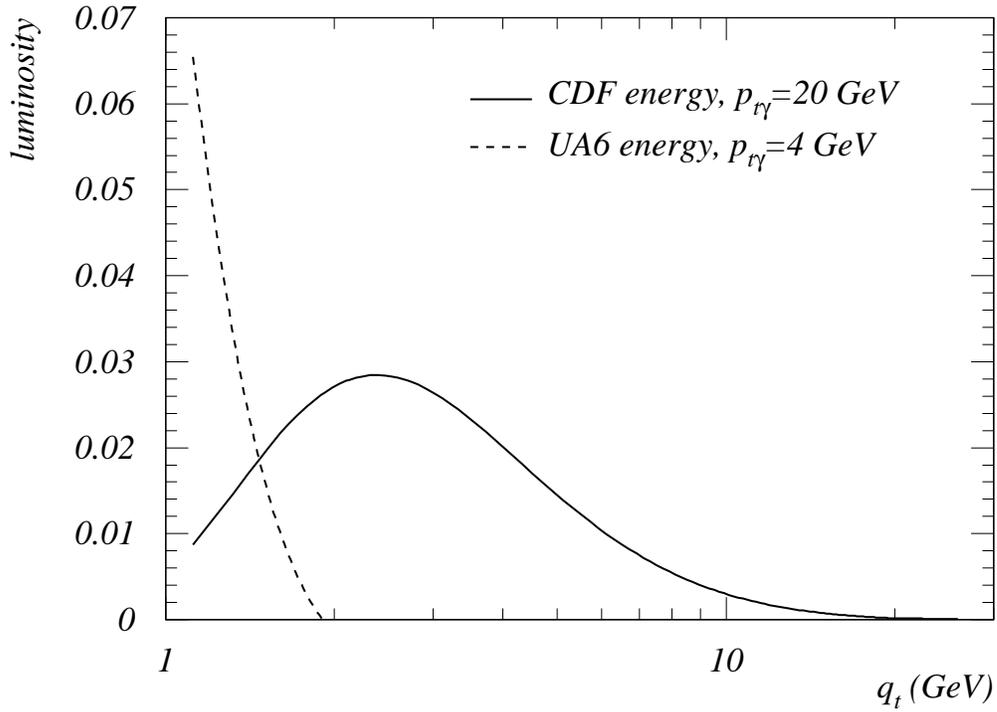}
\caption{The parton luminosity function ${\cal L}_{gq} (x_1, x_2, q_t)$ of (\ref{eq:a6})      
as a function of the incoming partonic $q_t$ for $p_{t \gamma} = 4$~GeV at the UA6    
energy of $\sqrt{s} = 24.3$~GeV and for $p_{t \gamma} = 20$~GeV at the Tevatron energy  
of $\sqrt{s} = 1800$~GeV.  In each case we take $x_1 = x_2 = x_T = 2p_{t  
\gamma}/\sqrt{s}$.  The hard scale $\mu$ is taken to be $p_{t \gamma}$.}
\label{fig2}
\end{figure}

It is informative to look at the form of the luminosity function ${\cal L}_{gq}$ as a function    
of the transverse momentum $q_t$ of the incoming partonic $(gq)$ system, before we    
consider the effects of $q_t$ on the $p_{t\gamma}$ spectrum of hadroproduced prompt    
photons.  Fig.~\ref{fig2} shows ${\cal L}_{gq}$ obtained from (\ref{eq:a6}), for values
 of $q_t$ in    
the perturbative domain $(q_t > q_0)$, at the energies of the UA6 and CDF experiments.  We    
use MRS(R2) partons \cite{MRSR} and take $q_0^2 = 1.25$~GeV$^2$.  At the higher  
Tevatron energy, $\sqrt{s} = 1.8$~TeV, we see that the $q_t$ distribution is extensive,  
whereas at the lower energy, $\sqrt{s} = 24.3$~GeV, the luminosity goes negative for  
sufficiently large $q_t$.  At this point we set ${\cal L} = 0$ for larger values of $q_t$.  The  
negative values are an artefact of the double logarithmic approximation on which  
(\ref{eq:a6}) is based; the derivative $\partial T_a/\partial \ln \lambda^2$ does not exactly  
cancel the second (virtual) term in the DGLAP equation (\ref{eq:a1}).  The use of the full  
treatment based on (\ref{eq:a3}) and (\ref{eq:xx}) would not suffer from this defect. 
      
We must, in addition, consider partonic transverse momentum in the domain $q_t < q_0$.  In    
this region we cannot use perturbative QCD and, moreover, we have no knowledge of the    
parton distributions at scales $\lambda \lapproxeq q_0$.  On the other hand a significant part    
of the total luminosity comes from the region $q_t < q_0$ for prompt photon hadroproduction    
at the lower energies, that is $\sqrt{s} \sim 30$~GeV.  We may quantify this since the {\it    
integrated} luminosity ${\cal L}_0$ coming from the region $q_t < q_0$ can be obtained    
directly from (\ref{eq:a6})     
\be     
\label{eq:a15}     
{\cal L}_0 \; = \; a (x_1, q_0^2) \: T_a (q_0, \mu) \: b (x_2, q_0^2) \: T_b (q_0, \mu).     
\ee     
Typical results are given in Table 1 in the relevant energy range.  We show the luminosity    
integrated over the perturbative domain $(q_t > q_0)$ together with ${\cal L}_0$ for given    
values of $x_1 = x_2$.  We have fixed the value of $x_1 = x_2 = 0.3$ in order to investigate  
the energy behaviour of the luminosity.  For fixed $x_1 = x_2$ and increasing $\sqrt{s}$, the  
corresponding value of $p_{t \gamma}$ increases and we see that the fraction of the  
luminosity coming from the non-perturbative region decreases rapidly.  At $\sqrt{s} =  
1800$~GeV we also give the luminosity at a more experimentally representative value, $x_1  
= x_2 = 0.03$, where it is much larger. 
\begin{table}[htb]   
\caption{The integrated luminosity from the perturbative $(q_t > q_0)$ and non-perturbative    
$(q_t < q_0)$ domains for typical values of $\sqrt{s}, x_1, x_2$.  We take $q_0^2 =    
1.25$~GeV$^2$ and the hard scale $\mu = p_{t \gamma}$.}   
\begin{center}   
\begin{tabular}{|c|c|c|c|c|} \hline   
$\sqrt{s}$ (GeV) & $x_1 = x_2$ & ${\cal L}_{gq} (q_t > q_0)$ & ${\cal L}_{gq} (q_t <    
q_0)$ & Fraction from \\   
& & & & $q_t < q_0$ \\ \hline   
25 & 0.3 & 0.062 & 0.160 & 0.72 \\   
50 & 0.3 & 0.095 & 0.048 & 0.34 \\   
100 & 0.3 & 0.0851 & 0.0103 & 0.11 \\   
500 & 0.3 & 0.0444 & 0.0001 & $2 \times 10^{-3}$ \\   
1800 & 0.3 & 0.0291 & $1.4 \times 10^{-6}$ & $5 \times 10^{-5}$ \\   
1800 & 0.03 & 1.16 & 0.0016 & $1 \times 10^{-3}$ \\ \hline   
\end{tabular}   
\end{center}   
\end{table}   
   
\begin{figure}\centering
\includegraphics{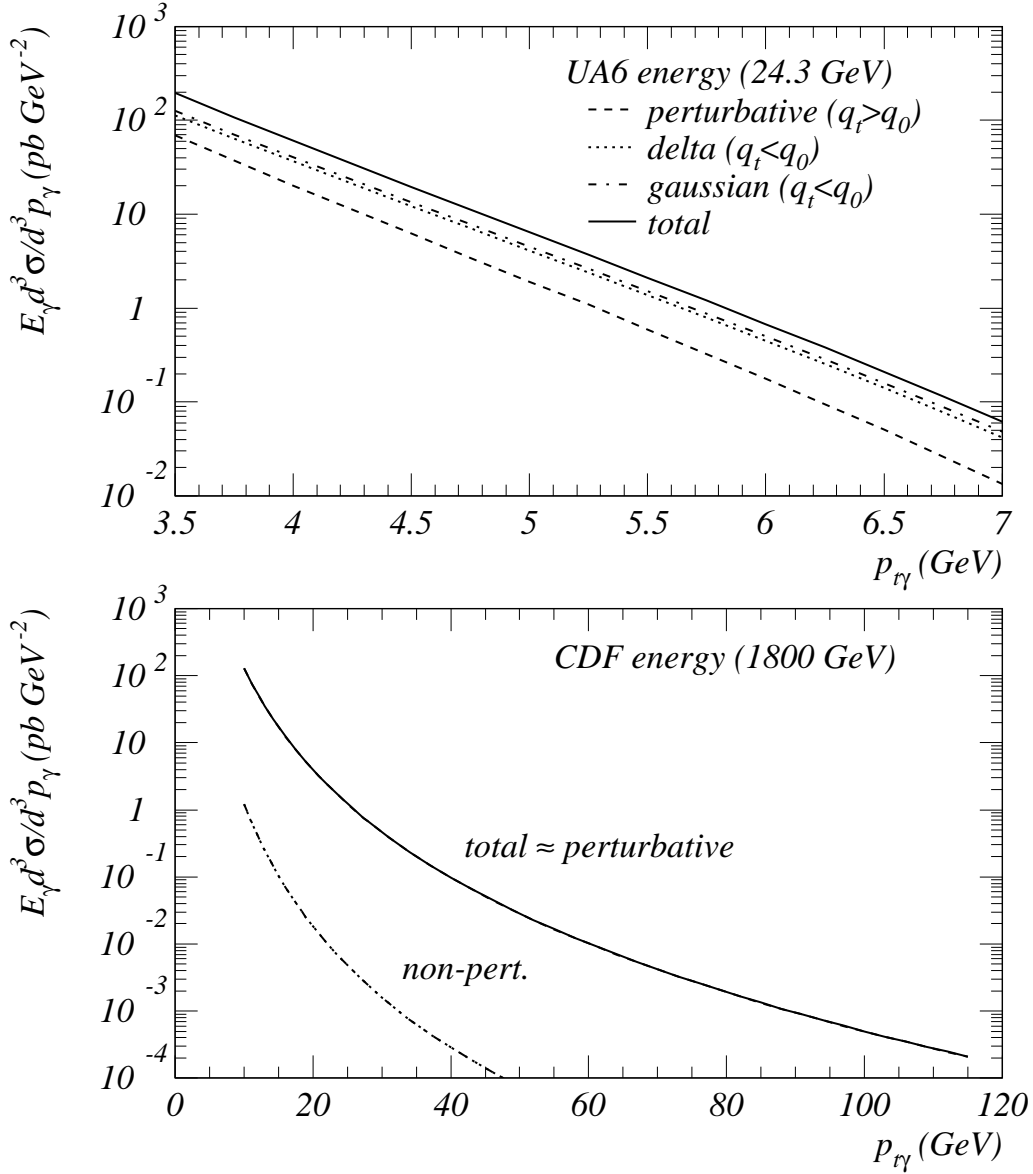}
\caption{The effect of partonic $q_t$ arising from the non-perturbative $(q_t < q_0)$,  
perturbative $(q_t > q_0)$, and total contributions to the $p_{t \gamma}$ spectrum of  
prompt photon production in $pp$ collisions at $\sqrt{s} = 24.3$~GeV and in $p\bar{p}$ 
collisions at 1800~GeV.  Two extreme hypotheses are used for the non-perturbative 
contribution, but the predictions are very similar (and in fact are indistinguishable on the 
$\sqrt{s} = 1800$~GeV plot).  We see that the non-perturbative contribution dominates at 
$\sqrt{s} = 24.3$~GeV, whereas the perturbative dominates at $\sqrt{s} = 1800$~GeV.  The 
hard scale $\mu$ is taken to be $p_{t \gamma}$ and $q_0^2 = 1.25$~GeV$^2$.}
\label{fig3}
\end{figure}

In order to investigate the effect of the partonic $q_t$ on the prompt photon $p_{t \gamma}$    
spectrum we need, not just the integrated luminosity ${\cal L}_0$, but also the $q_t$    
distribution in the non-perturbative region.  We take two extreme possibilities for the    
distribution in the region $q_t < q_0$.  First, we assume that there is no intrinsic transverse    
momentum in the non-perturbative region, and      
set     
\be     
\label{eq:a16}     
{\cal L} (q_t < q_0) \; = \; {\cal L}_0 \: \delta (q_t^2)     
\ee     
with ${\cal L}_0$ given by (\ref{eq:a15}).  Second, we assume a wide Gaussian      
distribution of partonic $q_t$,     
\be     
\label{eq:a17}     
{\cal L} (q_t < q_0) \; = \; {\cal L}_0 \: \frac{e^{-q_t^2/q_0^2}}{q_0^2 (1 - e^{-1})},     
\ee     
normalized to the region $q_t < q_0$.  From the results shown in Fig.~\ref{fig3} 
it turns out that the  
two extreme hypotheses, (\ref{eq:a16}) and (\ref{eq:a17}), for the non-perturbative $q_t$      
distribution lead to a similar $p_{t \gamma}$ spectrum.  Insight into why this result may be      
anticipated can be obtained from the following simple estimate.     
     
On average the prompt photon gets one half of the $q_t$ of the incoming partonic system.       
Thus the effect of the partonic $q_t$ on the hard subprocess cross section is     
\bea     
\label{eq:a18}     
\frac{d \hat{\sigma}}{d\hat{t}} \; \sim \; \frac{1}{p_t^{\prime 4}} & = & \frac{1}{\left      
|\mbox{\boldmath $p$}_{t \gamma} - \frac{1}{2} \mbox{\boldmath $q$}_t \right |^4}       
\nonumber \\     
& & \nonumber \\     
& \rightarrow & \frac{1}{p_{t \gamma}^4} \: \left [ 1 \: + \: \frac{3 (\mbox{\boldmath      
$p$}_{t \gamma} \cdot \mbox{\boldmath $q$}_t)^2}{p_{t \gamma}^4} \: - \: \frac{q_t^2}{2      
p_{t \gamma}^2} \: + \: \ldots \right ] \; \rightarrow \; \frac{1}{p_{t \gamma}^4} \: \left [ 1 \:      
+ \: \frac{q_t^2}{p_{t \gamma}^2} \: + \: \ldots \right ],     
\eea     
after the angular integration.  So the relative difference between the $q_t < q_0$ effects      
calculated using (\ref{eq:a16}) and (\ref{eq:a17}) is expected to be only about $q_0^2/2p_{t      
\gamma}^2$, since in the non-perturbative domain $\langle q_t^2 \rangle \lapproxeq q_0^2$.     
     
At this point one may ask why the traditional phenomenological treatments    
\cite{HUS,MRST} of the intrinsic partonic $q_t$ have found such a large effect.  The reason    
is that the full kinematics were not taken into account.  Instead of giving the Lorentz boost to    
the produced $\gamma +$parton system, the photon on its own received the total partonic  
$q_t$.     
To reproduce the same effect with the correct kinematics would require about twice as much    
partonic $q_t$.  Moreover strong-ordering of transverse momenta should be imposed which    
prevents the hard subprocess approaching the $\hat{t} \sim 0$ region.   
   
It is informative to discuss the results at the lowest (UA6) and the highest (Tevatron) energies      
separately.  At the lowest energy we see from Fig.~\ref{fig3} that the contribution from the 
perturbative $(q_t > q_0)$ region is relatively small.  Thus, even though it depends on the 
choice of the hard scale $\mu$, the ambiguity is not so important.  Moreover we note that the 
two extreme choices of the non-perturbative $q_t$ distribution, (\ref{eq:a16}) and 
(\ref{eq:a17}), give only a 20\% uncertainty in the $p_{t \gamma}$ spectrum.  That is 
partonic $q_t$ smearing has little effect on the $p_{t \gamma}$ spectrum, as may be 
anticipated from (\ref{eq:a18}).  However there are other related points to consider.  We see 
that the formalism requires that the parton densities are sampled at scale $q_0$, which 
enhances the cross section in comparison with the conventional leading order treatment in 
which the partons are sampled at the hard scale $\mu$.  The $q_t < q_0$ contribution 
depends on the scale $\mu$ through the survival probabilities $T_a$ and $T_b$ in ${\cal 
L}_0$ of (\ref{eq:a15}).  If, for illustration, we take the hard scale $\mu = cp_{t \gamma}$ 
then, to double $\log$ accuracy with fixed $\alpha_S$, we have     
\be     
\label{eq:a19}     
T_a \; \simeq \; \exp \: \left ( - \: \frac{C_a}{4 \pi} \: \alpha_S (cp_{t \gamma}) \: \ln^2 \: \left      
( \frac{cp_{t \gamma}}{q_0} \right ) \right ).     
\ee     
Thus an increase in scale, that is a larger $c$, makes $d\sigma/dp_{t \gamma}$ steeper and      
the cross section smaller.   The prediction for the prompt photon spectrum, $d\sigma/dp_{t  
\gamma}$, at the Tevatron energy is much more certain.  Here the contribution for the  
non-perturbative $q_t < q_0$ domain is negligible (see Fig.~\ref{fig3}), and so the calculation is more  
under control and is less scale dependent.  The difference between using (\ref{eq:a6}) and  
using an \lq\lq unsmeared\rq\rq~luminosity     
\be     
\label{eq:a20}     
{\cal L}_{ab} \; = \; \delta (q_t^2) \: a (x_1, \mu^2) \: b (x_2, \mu^2)     
\ee     
is not great.  The difference is due to the $q_t$ smearing itself and also due to the fact that in      
(\ref{eq:a6}) the partons are sampled at $q_t$.  This latter effect was not incorporated in      
the existing treatments of partonic smearing of the $p_{t \gamma}$ distribution.   

\begin{figure}\centering
\includegraphics{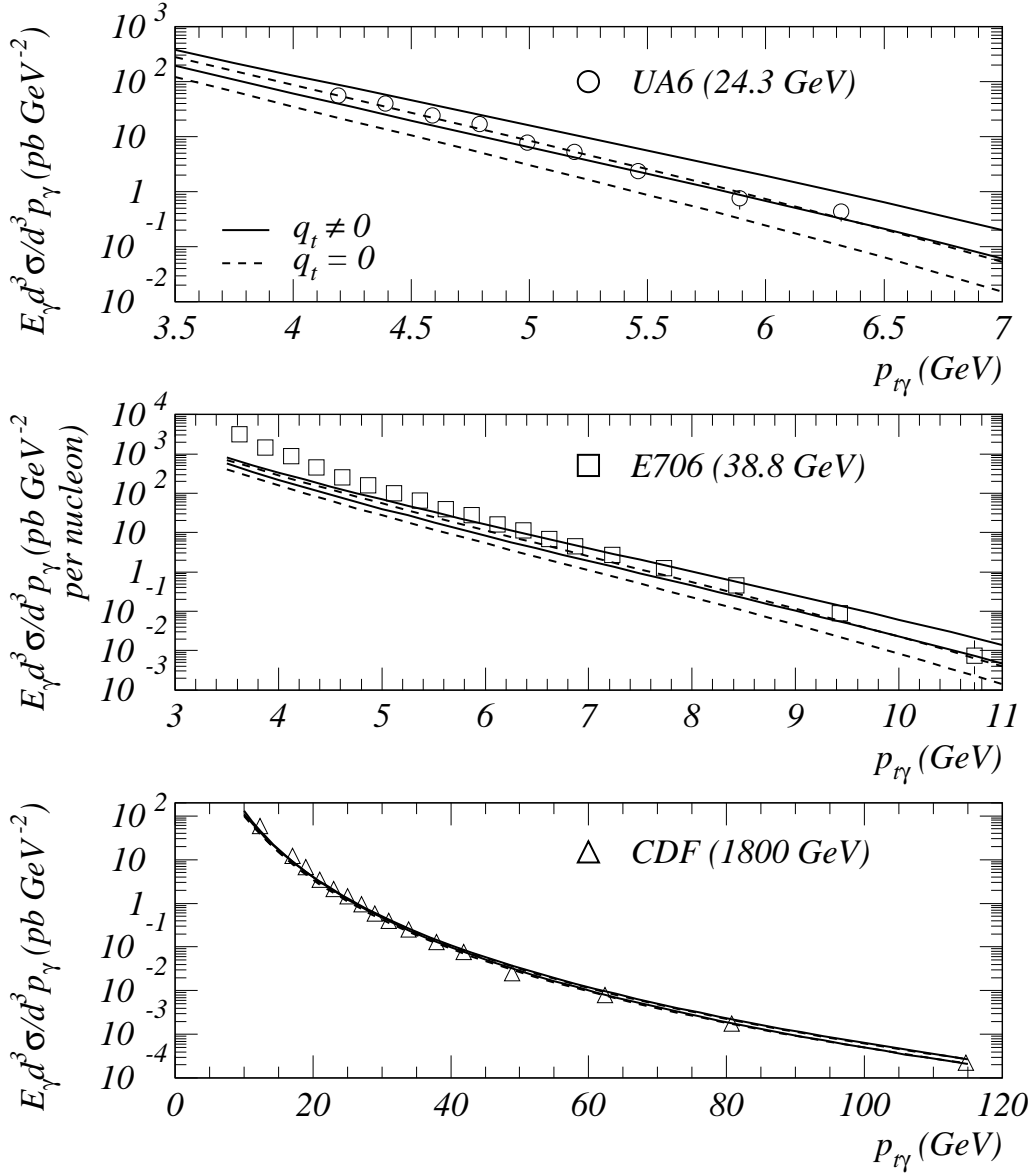}
\caption{The scale dependence of the predictions for production of prompt photons in  
$pp$ collisions at $\sqrt{s} = 24.3$~GeV, $p$Be collisions at $\sqrt{s} = 38.8$~GeV and  
$p\bar{p}$ collisions at $\sqrt{s} = 1.8$~TeV shown together with UA6 \cite{UA6}, E706  
\cite{E706} and CDF \cite{FNAL} data respectively.  The continuous curves are the  
predictions with the incoming partonic transverse momentum $q_t$ included, whereas the  
dashed curves correspond to the unsmeared $(q_t = 0)$ results in which the integrated partons  
are sampled at the hard scale $\mu$.  In each case the upper curve corresponds to the scale  
$\mu = p_{t \gamma}/2$, while the lower corresponds to $\mu = p_{t \gamma}$.}
\label{fig4}
\end{figure}

\begin{figure}\centering
\includegraphics{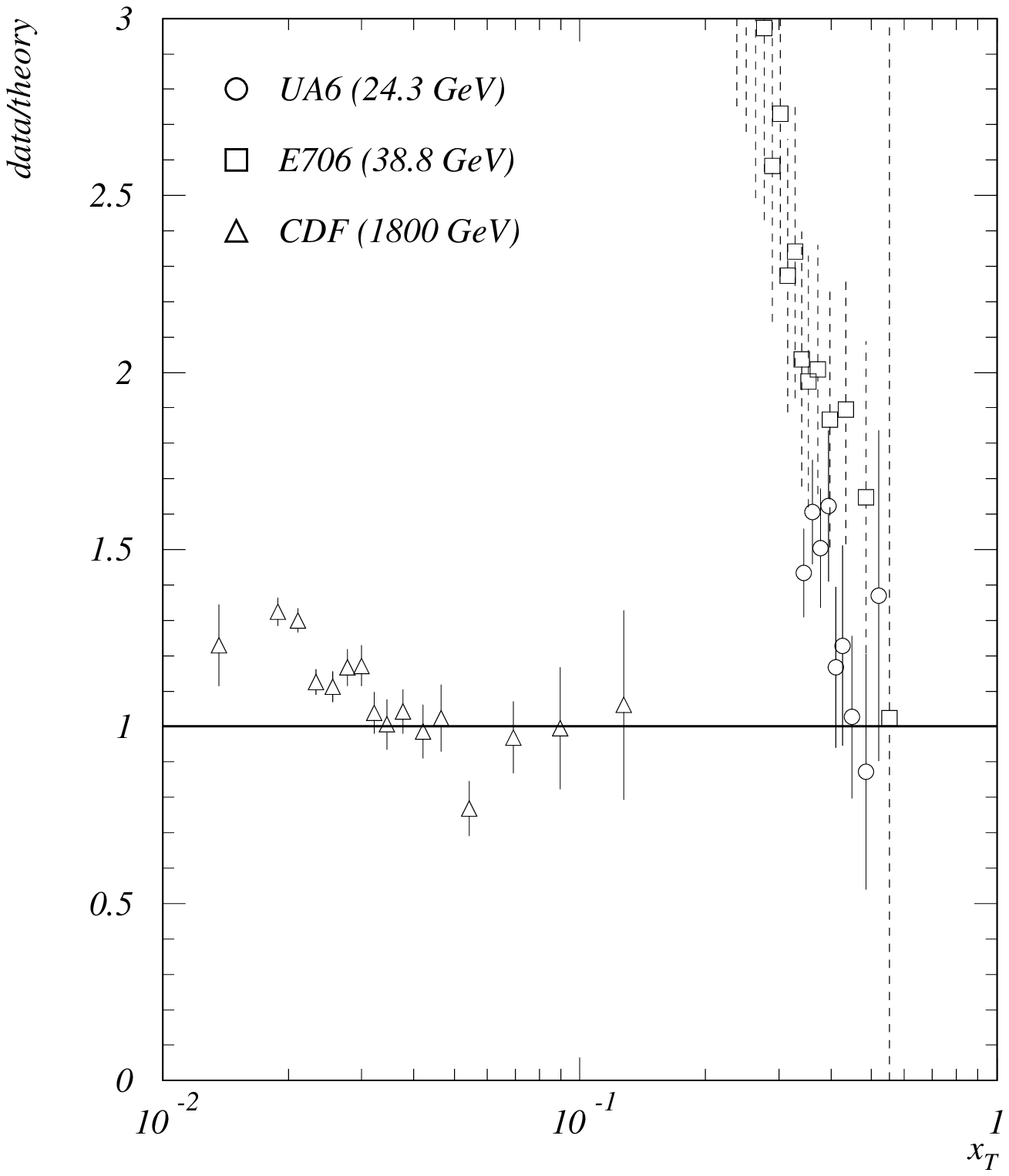}
\caption{A clearer comparison of the data also plotted in Fig.~\ref{fig4} with our
theoretical calculations at the scale $\mu = p_{t \gamma}$; we use 
$x_{T}\equiv 2p_{t\gamma}/\sqrt{s}$ as the transverse momentum variable.}
\label{fig5}
\end{figure}

To demonstrate the sensitivity to the choice of the hard scale, in Fig.~\ref{fig4}
 we compare the 
results for $\mu = p_{t \gamma}/2$ with those for $\mu = p_{t \gamma}$ for three typical 
energies at which experimental measurements exist.  A third choice of the hard scale, $\mu = 
M/2$, gives similar results to those obtained with $\mu = p_{t \gamma}$.  For comparison 
the unsmeared cross sections are shown by dashed curves.  Very little of the difference 
between the continuous $(q_t \neq 0)$ and the dashed $(q_t = 0)$ curves is directly 
attributable to the introduction of the partonic transverse momentum $q_t$.  As mentioned 
above, the main difference is due to the proper inclusion of the parton densities, which are 
sampled at scale $q_t$ (with $q_t > q_0$) in contrast to the conventional unsmeared result in 
which they are sampled at the hard scale $\mu$.  Of course in a realistic fit to the data we 
would need to implement the more precise form (\ref{eq:a3}) of the unintegrated parton 
densities, as well as computing the next-to-leading order expression for the hard subprocess 
cross sections.  However for the largest energy, $\sqrt{s} = 1.8$~TeV, where the $p_{t 
\gamma}$ is large and leading order is a good approximation, the theoretical uncertainties are 
seen to be small\footnote{The difference between the predictions for $\mu = p_{t \gamma}$ 
and $\mu = p_{t \gamma}/2$ may be considered as an estimate of the size of the next-to-
leading order contribution.}.  In fact the $\mu = p_{t \gamma}$ prediction gives a reasonable 
description of the CDF data, particularly at the larger values of $p_{t\gamma}$, see 
Fig.~\ref{fig5}.  Figs.~\ref{fig4} and \ref{fig5} also show that, in common with other 
analyses\footnote{For example, see the (unsmeared) results in Refs.~\cite{HUSTON,AUR,HUS,MRST}.},
there is a large disagreement between the predictions and the $p_{t\gamma}$ shape of the
E706 data.
     
\section{Conclusions}     
This paper should not be regarded as an attempt to better fit data, such as the $p_{t      
\gamma}$ spectrum of prompt photon hadroproduction, but rather to illustrate the appropriate      
formalism that should be used to incorporate the transverse momenta of the incoming partons.       
The crucial point is that, in general, according to the $k_{t}$-factorization prescription
\cite{KT}, cross sections should be calculated in terms of parton      
distributions, $f_a (x, k_t^2, \mu^2)$, {\it unintegrated} over the transverse momentum.       
Such distributions can be constructed from the conventional parton densities, $a (x,      
\lambda^2)$, as described in Section~2.     
     
Prompt photon hadroproduction is an ideal example to illustrate the necessity of using      
unintegrated parton distributions.  For simplicity we chose to work to double logarithm      
accuracy, rather than the full treatment presented in Section~2.  Contrary to the existing      
phenomenological treatments of the intrinsic transverse momenta of the partons, we find that      
the $q_t$ effects give themselves only a small modification to the $d\sigma/dp_{t \gamma}$      
spectrum.  Of course there could have been a large effect if $q_t \sim p_{t \gamma}$ which  
would have allowed the singularity of the hard subprocess amplitude to be approached.   
However the strong ordering of transverse momenta must be incorporated in the formalism,  
via $\theta (| \hat{t} |  - q_t^2)$, which prevents this happening.  In addition, our study reveals  
that the $q_t$ of the incoming partonic system should be correlated with the scale at which  
the parton densities are sampled.  This fact affects both the shape and the normalization of the  
predictions, especially at the lower energies.     
     
An interesting result of our study of the parton luminosity is that, at low energy where the    
contribution from the $q_t > q_0$ region is small, the factorization procedure is more direct    
and physically transparent.  Recall that we have to introduce a factorization scale to separate    
the hard subprocess interaction (which is calculated perturbatively) from the universal parton    
distributions which originate from large distances (that is, from the confinement region).     
These incoming partons starting at the input scale $q_0$ participate in the perturbative    
evolution which changes the original distributions.  However for larger $x$ (that is for    
the lower energies) we sample the parton densities directly at the low input scale $q_0$, see    
Fig.~\ref{fig2} and (\ref{eq:a15}).  From the form of the luminosity we see that essentially no new    
partons appear in the evolution to larger values of $q_t$.  The only perturbative effect is the    
introduction of the survival probability, $T_a$, which has a simple physical interpretation      
and does not alter the kinematics of the partons sampled.  This means that the low energy    
data\footnote{For example, UA6 data at $p_{t \gamma} = 5$~GeV.} measure the input distributions 
directly.  On the other hand at the  
smaller values\footnote{For example, Tevatron data at $p_{t \gamma} = 40$~GeV.} of $x$  
the perturbative contribution $q_t > q_0$ dominates.  In this case the partons evolve from the  
input scale $q_0$ and are finally sampled at the hard factorization scale $\mu \sim p_{t  
\gamma}$.  The evolution of the integrated parton densities in the conventional approach  
modifies the $x$ behaviour of the distributions assuming that the transverse momenta $k_t$  
of the partons is small.  The formalism that is presented here embodies the modification of  
the $k_t$ distributions of the partons during the evolution, as well as the modification of their 
$x$ dependence.  That is we are able to determine, perturbatively, the $q_t = | 
\mbox{\boldmath $k$}_{1t} + \mbox{\boldmath $k$}_{2t} |$ distribution of the incoming 
partonic system, which leaves no room for additional phenomenological $q_t$ smearing.     
   
In summary, we have presented the formalism to construct the parton distributions  
unintegrated over the parton transverse momenta, $f_a (x, k_t^2, \mu^2)$, from the known    
conventional parton distributions.  These distributions should be used to calculate cross    
sections initiated by hadrons.  We applied the formalism to prompt photon hadroproduction in    
the double logarithm approximation.  In this way we are able to quantify the effects of the    
partonic transverse momentum, $q_t$, on the transverse momentum, $p_{t \gamma}$,  
spectrum of the produced photons.  We find that the correct treatment requires that the  
integrated partons are sampled at the scale $q_t$.  Clearly this is only meaningful in the  
perturbative domain $q_t > q_0$, but we are able to show that the uncertainty due to the  
contribution from the non-perturbative domain is only of relative size of the order of 
$q_0^2/2p_{t\gamma}^2$.   \\ 
 
\noindent {\large \bf Acknowledgements} 
 
We thank Dick Roberts for discussions.  One of us (MGR) thanks the Royal Society and the  
Russian Fund for Fundamental Research (98-02-17629) for support.  The work was also  
supported by the UK Particle Physics and Astronomy Research Council and the EU 
Framework TMR programme, contract FMRX-CT98-0194.


\newpage      
\begin{thebibliography}{xx}   
\bibitem{BFKL} E.A. Kuraev, L.N. Lipatov and V.S. Fadin, Sov. Phys. JETP {\bf 45} (1977)  
199; \\ 
Ya.Ya. Balitzkij and L.N. Lipatov, Sov. J. Nucl. Phys. {\bf 28} (1978) 822; \\
L.V. Gribov, E.M. Levin and M.G. Ryskin, Phys. Rep. {\bf 100} (1983) 1.    
\bibitem{DDT} Yu.L. Dokshitzer, D.I. Dyakanov and S.I. Troyan, Phys. Rep. {\bf 58} (1980)      
269.
\bibitem{KT} S. Catani, M. Ciafaloni and F. Hautmann, Phys. Lett. {\bf B242} (1990) 97; \\
Nucl. Phys. {\bf B366} (1991) 135; \\
J.C. Collins and R.K. Ellis, Nucl. Phys. {\bf B360} (1991) 3; \\
G. Marchesini and B.R. Webber, Nucl. Phys. {\bf B386} (1992) 215; \\
S. Catani and F. Hautmann, Nucl. Phys. {\bf B427} (1994) 475.    
\bibitem{HUSTON} J. Huston et al., Phys. Rev. {\bf D51} (1995) 6139.     
\bibitem{AUR} P. Aurenche, M. Fontannaz, J.Ph. Guillet, B. Kniehl, E. Pilon and M.      
Werlen, Eur. Phys. J. {\bf C9} (1999) 107.   
\bibitem{HUS} L. Apanasevich et al., Phys. Rev. {\bf D59} (1999) 074007.     
\bibitem{MRST} A.D. Martin, R.G. Roberts, W.J. Stirling and R.S. Thorne, Eur. Phys. J.      
{\bf C4} (1998) 463.     
\bibitem{UA6} UA6 collaboration, G. Ballocchi et al., Phys. Lett. {\bf B436} (1998) 222.     
\bibitem{E706} E706 collaboration, L. Apanasevich et al., Phys. Rev. Lett. {\bf 81} (1998)      
2642.     
\bibitem{FNAL} CDF collaboration, F. Abe et al., Phys. Rev. Lett. {\bf 73} (1994) 2662.     
\bibitem{CATANI} S. Catani, M. Mangano, P. Nason, C. Oleari and W. Vogelsang, JHEP  
9903:025 (1999). 
\bibitem{LL} H.-L. Lai and H.-N. Li, Phys. Rev. {\bf D58} (1998) 114020.     
\bibitem{CTEQ} CTEQ collaboration:  H.-L. Lai et al., {\tt hep-ph/9903282}. 
\bibitem{MW} G. Marchesini and B.R. Webber, Nucl. Phys. {\bf B310} (1988) 461.      
\bibitem{MRSR} A.D. Martin, R.G. Roberts and W.J. Stirling, Phys. Lett. {\bf B387} (1996)  
419. 
\end{thebibliography}
\end{document}